\documentstyle[aps,twocolumn,epsf]{revtex}
\def\ave#1{\langle #1\rangle}
\newcommand{\ve}[1]{{\vec #1}}

\newcommand{\Ord}[1]{{\cal O}\left(#1\right)}
\newcommand{\bra}[1]{\langle #1|}
\newcommand{\ket}[1]{|#1\rangle}
\newcommand{\braket}[2]{\langle #1|#2\rangle}
\newcommand{\half}{\textstyle{\frac{1}{2}}}
\begin{document}
\title{Time evolution of a quantum many-body system:
transition from integrability to ergodicity in thermodynamic limit}
\author{Toma\v z Prosen}
\address{Physics Department, Faculty of Mathematics and Physics,
University of Ljubljana, Jadranska 19, 1111 Ljubljana, Slovenia
}  
\date{\today}
\draft
\maketitle
\begin{abstract}
Numerical evidence is given for non-ergodic (non-mixing) behavior, exhibiting 
ideal transport, of a simple non-integrable many-body quantum system in the 
thermodynamic limit, namely kicked $t-V$ model of 
spinless fermions on a ring. However, for sufficiently large kick parameters 
$t$ and $V$ we recover quantum ergodicity, and normal transport, which can be 
described by random matrix theory.
\end{abstract}

\vfill

A simple question is addressed here: ``Do mixed quantum many-body systems,
which are neither integrable nor ergodic, exist in the thermodynamic limit?''
While it is clear that integrable systems are rather exceptional it is an
important open question whether a finite generic perturbation of an 
integrable system becomes ergodic or not in the thermodynamic limit (TL),
{\em size} $\rightarrow\infty$ and {\em fixed density}. 
It is known that local statistical properties of quantum systems with 
few {\em degrees of freedom} whose classical limit is completely chaotic ---
{\em ergodic} are universally described by {\em random matrix theory}, 
while in the other extreme case of {\em integrable} systems
Poissonian statistics may typically be applied \cite{qchaos,H91}. 
This statement has recently
been verified numerically also for integrable and {\em strongly}
non-integrable many-body systems of interacting fermions \cite{fRMT}
which do not have classical limit.

Having lost the reference to classical dynamics 
we resort to the definition of {\em quantum ergodicity} (also termed 
{\em quantum mixing}) \cite{JLP96} as the decay of time-correlations 
$\ave{A(\tau)B(0)}-\ave{A}\ave{B}$ of any pair of quantum observables $A$ and 
$B$ in TL, taking the time-limit $\tau\rightarrow\infty$ in the end. 
In \cite{JLP96} many-body system of interacting bosons has been studied and
it has been shown that quantum ergodicity corresponds to strongly chaotic
(classically ergodic) dynamics of associated non-linear mean-field equations.
As a consequence of linear response theory, quantum ergodicity also
implies normal transport and {\em finite} transport coefficients
(such as dc electrical conductivity).
On the other hand, {\em integrable systems}, which are solvable by Bethe
ansatz or quantum inverse scattering, are characterized by (infinitely many)
conservation laws and are thus {\em non-ergodic}.
It has been pointed out recently \cite{Prelovsek} that integrability
implies nonvanishing stiffness, i.e. ideal conductance with infinite 
transport coefficients (or ideal insulating state).
As we argue below, any deviation from quantum ergodicity generically implies
nonvanishing long-time current auto-correlation and therefore infinite
transport coefficient. Since generic non-integrable systems of finite size
(number of degrees of freedom) are non-ergodic (obeying {\em mixed}
statistics smoothly interpolating from Poissonian to random matrix
results) it is thus an
important question if and when such non-ergodicity can survive TL.

In this Letter we introduce a family of simple many-body systems smoothly
interpolating between integrable and ergodic regime, namely 
{\em kicked t-V model} (KtV) of spinless fermions with periodically switched 
nearest neighbor-interaction on a 1-dim lattice of size $L$ and periodic 
b.c. $L\equiv 0$, with time-dependent hamiltonian
\begin{equation}
H(\tau) = \sum_{j=0}^{L-1} \left[-\half t(c^\dagger_j c_{j+1} + h.c.) +
\delta_p(\tau) V n_j n_{j+1}\right],
\label{eq:hamiltonian}
\end{equation}
and give numerical evidence for
the existence of {\em mixed regime} (reader should not confuse it 
with {\em mixing}) in TL by direct simulation of the time
evolution. $c_j^\dagger,c_j,n_j$ are fermionic creation, annihilation and
number operators, respectively, and 
$\delta_p(\tau)=\sum_{m=-\infty}^\infty \delta(\tau-m)$. 
Deviations from quantum ergodicity (or mixing) 
are characterized by several different quantities as described below.
\\\\
KtV model (\ref{eq:hamiltonian}) is a many-body analogue of popular 1-dim
non-integrable kicked systems \cite{H91} 
such as e.g. kicked rotor: its evoulution (Floquet) operator over one 
period $U = \hat{T}\exp(-i\int_{0^+}^{1^+} d\tau H(\tau))$ ($\hbar=1$),
factorizes into the product of kinetic and potential part
\begin{equation}
U = \exp\left(-iV\sum_{j=0}^{L-1} n_j n_{j+1}\right)
\exp\left(it\sum_{k=0}^{L-1} \cos(s k + \phi)\tilde{n}_k\right)
\label{eq:map}
\end{equation}
where $s=2\pi/L$. Flux parameter $\phi$ is used in order to introduce
a {\em current operator}
$J = (i/t) U^\dagger \partial_\phi U\vert_{\phi=0} =
\sum_{k=0}^{L-1} \sin(s k)\tilde{n}_k$,
elsewhere we put $\phi:=0$.
Tilde denotes the operators which refer to {\em momentum} variable $k$,
$\tilde{c}_k = L^{-1/2}\sum_{j=0}^{L-1} \exp(i s jk) c_j,\;
\tilde{n}_k = \tilde{c}_k^\dagger \tilde{c}_k$.
KtV model is integrable if either $t=0$, or $V=0 \pmod{2\pi}$, or
$tV\rightarrow 0$ and $t/V$ finite (continuous time t-V model, see e.g. 
\cite{Emery}), while for $t\sim V\sim 1$ it is expected to be 
non-integrable, either {\em quantum ergodic} or {\em mixed}.

We expect that unitary many-body quantum maps, such as (\ref{eq:map}), 
also mimick the dynamics of generic {\em autonomous} quantum many-body
systems on the energy shell in a similar way as 1-dim 
quantum maps describe (quantum) Poincar\' e 
sections of 2-dim quantum (chaotic) systems (see e.g. \cite{P96}).

The total number of particles $N=\sum_j n_j$ is conserved, so the map $U$ 
acts over Hilbert (Fock) space ${\cal H}$ of dimension 
${\cal N} = {L \choose N}$. The dynamics of a given initial 
many body state $\ket{\psi(0)}$, which is an
iteration of the map 
$\ket{\psi(m)} = U\ket{\psi(m-1)} = U^m\ket{\psi(0)}$,
can be performed most efficiently by observing that the kinetic part 
$U_T$ is diagonal in the momentum basis $\ket{\ve{k}} =
\tilde{c}^\dagger_{k_1}\cdots\tilde{c}^\dagger_{k_N}\ket{0},
k_1 < \ldots < k_N$ while the potential part $U_V$ is diagonal in the
position basis 
$\ket{\ve{j}} = c^\dagger_{j_1}\cdots c^\dagger_{j_N}\ket{0}, 
j_1 < \ldots < j_N$. The transformation between the two,
$F_{\vec{j}\vec{k}} = \braket{\vec{j}}{\vec{k}}$,
is an antisymmetrized N-dim discrete Fourier transformation (DFT) on
L-sites which has been efficiently coded in
$\sim {\cal N}\log_2{\cal N}$ floating point operations (FPO)
by factorizing $L-$site DFT to the product of $\Ord{L\log_2 L}$ 
2-site transformations parametrized with $2\times 2$ sub-matrices
$(\alpha,\beta;\gamma,\delta)_{j j^\prime}$, which are successively 
applied to creation operators, 
$(c^\dagger_j,c^\dagger_{j^\prime})\leftarrow
(\alpha c^\dagger_j + \beta c^\dagger_{j^\prime},
 \gamma c^\dagger_j + \delta c^\dagger_{j^\prime})$,
in all slater determinants $\Pi_n c^\dagger_{j_n}\ket{0}$ 
which contain a particle at sites $j$ or $j^\prime$.
Our algorithm (fermionic FFT) requires almost no
extra storage apart from a vector of ${\cal N}$ c-numbers and works for 
lattices of sizes $L=2^p,10,12,15,20,24,30,40$.
Therefore, the map (\ref{eq:map}) is iterated on a vector
$\psi_{\vec{k}}(m) = \braket{\ve{k}}{\psi(m)}$, using the matrix composition
$U = F^* U_V F U_T $ in roughly $2{\cal N}\log_2 {\cal N}$ FPO per time step
which is by far superior to complete diagonalization techniques 
($\Ord{{\cal N}^3}$ FPO), even for {\em long} time scales $m=\Ord{\cal N}$ 
when quantum dynamics becomes quasiperiodic due to discreteness of the 
spectrum of $U$.

\begin{figure}[htbp]
\begin{center}
\leavevmode
\epsfxsize=3.6in
\epsfbox{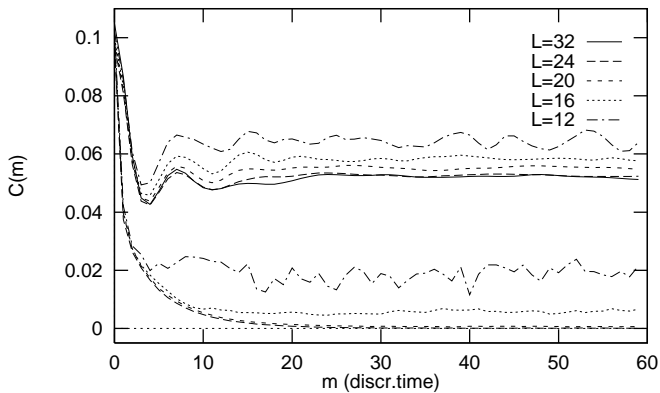}
\end{center}
\caption{Current autocorrelation function $C_J(m)$ against discrete time 
$m$ for quantum ergodic ($t=V=4$, lower set of curves for various sizes 
$L$) and mixed regime ($t=V=1$, upper set of curves) with
density $\rho=\frac{1}{4}$. Averaging over entire Fock space is performed, 
${\cal N}^\prime={\cal N}$, for $L\le 20$, whereas random samples of 
${\cal N}^\prime=12000$, and ${\cal N}^\prime=160$ initial states have 
been used for $L=24$, and $L=32$, respectively.}
\label{fig:1}
\end{figure}

\noindent
First we consider decay of current autocorrelation function
$C_J(m) = (1/L)\ave{J(m)J(0)}$ where $J(m) = U^{\dagger m} J U^m$,
and $\ave{.}=(1/{\cal N})Tr(.)$ is a `microcanonical
average'. Note that $J$ is diagonal in the momentum basis
$J\ket{\vec{k}} = J_{\vec{k}}\ket{\vec{k}}$, and $\ave{J} = 0$.
So $C_J(m)$ can be evaluated by means of time-evolution
of momentum initial states $\ket{\psi(0)}=\ket{\vec{k}^\prime}$
\begin{equation}
C_J(m)=\frac{1}{L\cal N^\prime}\sum^\prime_{\vec{k}^\prime} 
J_{\vec{k}^\prime}
\sum_{\vec{k}} J_{\vec{k}}\; p_{\vec{k}\vec{k}^\prime}(m)
\label{eq:corr}
\end{equation}
where $p_{\vec{k}\vec{k}^\prime}(m)=|\braket{\vec{k}}{\psi(m)}|^2=
|\bra{\vec{k}}U^m\ket{\vec{k}^\prime}|^2$.
For large sizes $L$, a smaller but uniformly random sample of
${\cal N}^\prime$ initial states $\ket{\vec{k}^\prime}$,
$1 \ll {\cal N}^\prime \ll {\cal N}$, is used in order to save computer
time. Direct computation of $C_J(m)$ for $m\le M$
can be performed in $\sim (2M{\cal N} {\cal N\,}^\prime /L)\log_2 {\cal N}$ 
FPO, since due to translational symmetry one can simultaneously 
simulate the dynamics of $L$ different states 
with different values of the conserved
total momentum $K=\sum_n k^\prime_n\pmod{L}$. 
Using the eigenphases $\eta_n$ and eigenstates $\ket{n}$
of evolution operator $U$, 
$U\ket{n}=e^{-i\eta_n}\ket{n}$,$n=1\ldots{\cal N}$, 
one can write {\em dissipative} dc conductivity of such a kicked 
system $\sigma := \sum_{n=1}^{\cal N} 
(\partial_\phi \eta_n)^2  \approx 
C_J(0) + 2\sum_{m=1}^{{\cal N}/2} C_J(m)$.
Note that $\partial_\phi\eta_n=\bra{n}J\ket{n}$.

\begin{figure}[htbp]
\begin{center}
\leavevmode
\epsfxsize=3.6in
\epsfbox{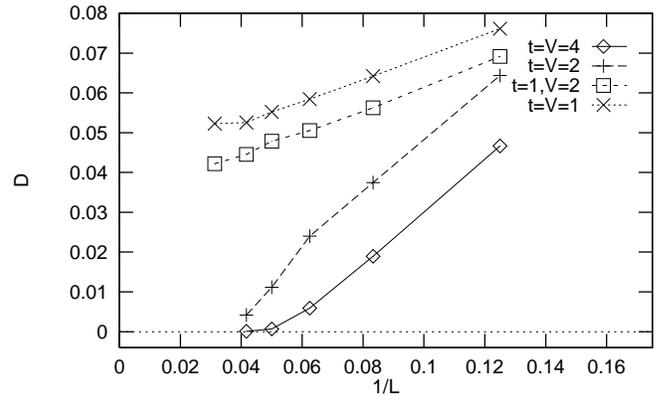}
\end{center}
\caption{Stiffness $D_J$ vs. $1/L$
at constant density $\rho=\frac{1}{4}$ and for different 
values of control parameters in ergodic, $t=V=4$ 
and $t=V=2$, and mixed, $t=1,V=2$ and $t=V=1$, regime.
Other parameters are the same as in fig.1.}
\label{fig:2}
\end{figure}

In fig.1 we present numerical computation of correlation 
function $C_J(m)$ for parameters $t=V=1$ and $t=V=4$, 
for various sizes $L$, but at fixed density $\rho=N/L=\frac{1}{4}$. 
Quite generally, $C_J(m)$ exhibits fast relaxation on a 
time scale $M^*$ which is typically small, $M^* \sim 10$, 
and roughly independent of $L$, and afterwards it 
fluctuates around averaged limiting value, the {\em stiffness}
\begin{equation}
D_J = \lim_{M\rightarrow\infty}\frac{1}{M}\sum_{m=1}^M C_J(m)
\label{eq:stif}
\end{equation}
where the strength of fluctuations decreases with incresing size $L$.
Note again that TL $L \rightarrow\infty$ should be taken prior
to the time-limit, $\lim_{M\rightarrow\infty}(1/M)\sum_{m=1}^M (.)$,
which is for systems of finite size $L$ here and below estimated 
numerically as $(1/M^\prime)\sum_{m=M^\prime+1}^{2M^\prime}(.)$ 
with sufficiently large but fixed
averaging time $M^\prime > M^*$; we take $M^\prime = 30$.
If the system is quantum ergodic (case $t=V=4$ of fig.1), 
$D_J$ goes to zero and $\sigma$ remains finite as 
$L\rightarrow\infty$ (${\cal N}\rightarrow\infty$) 
and $\rho=N/L$ fixed, whereas in the other case 
($t=V=1$ of fig.1) $D_J$ remains well above zero as we 
approach TL whereas conductivity $\sigma$ diverges\cite{footn}.  
In fig.2 we have analyzed the scaling of $D_J$ with $1/L$.
For large values of parameters, say $t=V=4$, $D_J$ is practically 
zero already for $L\approx 20$, while for smaller (but not small) 
control parmeters ($t,V$), 
$D_J \approx D^\infty_J + \beta/L$ where $D^\infty_J > 0$.
In {\em close-to-critical} case $t=V=2$, we find larger 
correlation time $M^*\sim 10^2$, and hence use longer 
averaging time $M^\prime = 200$.
In fig.3 we illustrate an {\em ideal transport} 
for $t\sim V\sim 1$ by plotting a {\em persistent current} 
$J^p_{\vec{k}^\prime}=\lim_{M\rightarrow\infty}(1/M)
\sum_{m=1}^M \bra{\vec{k}^\prime}J(m)\ket{\vec{k}^\prime}$
vs. the initial current $J_{\vec{k}^\prime}$. 
The normal transport in the ergodic
regime $t=V=4$ is characterized by $J^p_{\vec{k}^\prime}=0$, 
while for $t\sim V\sim 1$ we find ideal transport with the 
persistent current being proportional to the initial current,
$J^p_{\vec{k}^\prime} = \alpha J_{\vec{k}^\prime}$.
Proportionality constant $\alpha$ can be computed from 
(\ref{eq:stif})
$D_J = (1/L)\ave{J_{\vec{k}^\prime}J^p_{\vec{k}^\prime}} = 
(\alpha/L)\ave{J^2}$, so $\alpha = 2 D_J/[\rho(1-\rho)]$, 
where $\ave{J^2}$ is given below (\ref{eq:mc}).

\begin{figure}[htbp]
\begin{center}
\leavevmode
\epsfxsize=3.6in
\epsfbox{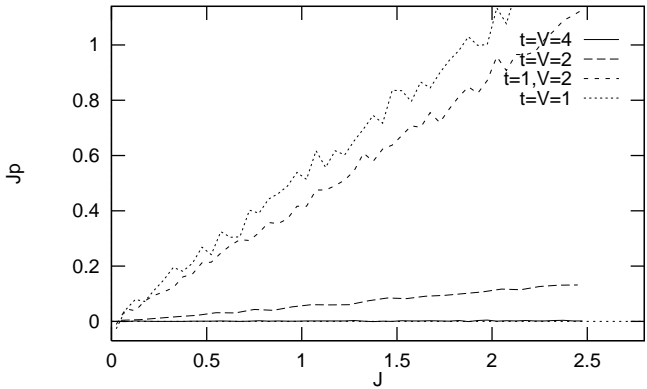}
\end{center}
\caption{Persistent current $J^p_{\vec{k}}$ against initial 
current $J_{\vec{k}}$ (averaged over bins of size 
$\Delta J=0.005$) in the ergodic $t=V=4$, (nearly) 
ergodic $t=V=2$, and mixed regime, $t=V=1$ and $t=1,V=2$. 
In all cases, $L=24$ and $\rho=\frac{1}{4}$.}
\label{fig:3}
\end{figure}

\noindent
Due to translational symmetry the total momentum 
$K=\sum_k k \tilde{n}_k\pmod{L}$ is the
only conserved quantity (appart from $N$ and parity), 
so the evolution of initial momentum state
$\ket{\vec{k}^\prime}$ takes place in 
${\cal N}_K \approx {\cal N}/L$ dim.
subspace ${\cal H}_K$, spanned by $\ket{\vec{k}}$ with 
$K=|\vec{k}|:=\sum_n k_n$.
Starting with a momentum state $\ket{\vec{k}^\prime}$,
the number of `excited' states $\ket{\vec{k}}$ after time $m$ is
characterized by information entropy \cite{I89} 
(see also \cite{P96}) as $\exp(-\sum_{\vec{k}} 
p_{\vec{k}\vec{k}^\prime}(m)\log p_{\vec{k}\vec{k}^\prime}(m))$.
Averaging the entropy over uniformly random sample of 
${\cal N}^\prime$ initial states $\ket{\vec{k}^\prime}$ we define 
{\em relative localization dimension} in Fock space as a measure
of quantum ergodicity.
$$
R(m) = \frac{L}{\cal N}\exp\left(-\frac{1}{{\cal N}^\prime}
\sum\limits^\prime_{\vec{k}^\prime}
\sum\limits_{\vec{k}} p_{\vec{k}\vec{k}^\prime}(m)\log
p_{\vec{k}\vec{k}^\prime}(m)\right).
$$
Again similar behavior is found numerically for $R(m)$ as for 
$C_J(m)$ (the two quantities can be computed simultaneously at 
no extra cost), namely it typically saturates within the same 
(short) correlation time $M^*$ to a roughly constant value 
$\bar{R} = \lim_{M\rightarrow\infty}(1/M)\sum_{m=1}^M R(m)$

\begin{figure}[htbp]
\begin{center}
\leavevmode
\epsfxsize=3.6in
\epsfbox{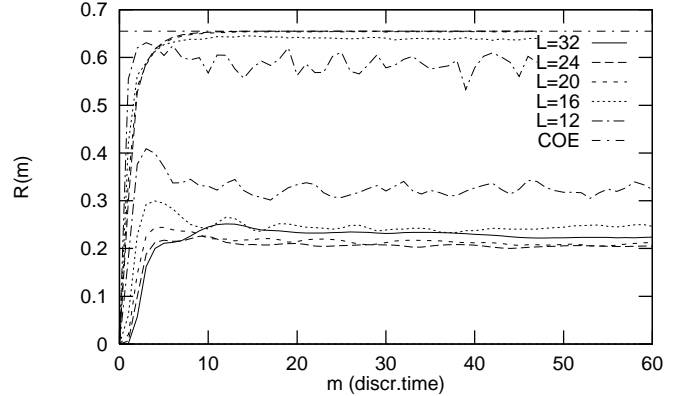}
\end{center}
\caption{Relative localization dimension in Fock space, 
$R(m)$ for data of fig.1.}
\label{fig:4}
\end{figure}
\begin{figure}[htbp]
\begin{center}
\leavevmode
\epsfxsize=3.6in
\epsfbox{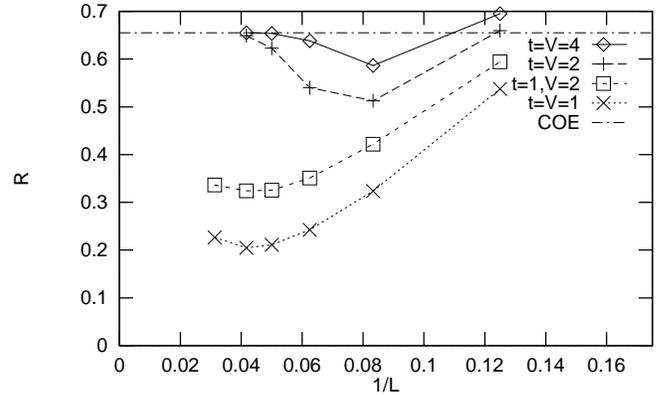}
\end{center}
\caption{Limiting rel.loc.dim. $\bar{R}$ vs. $1/L$ for data of fig.2.}
\label{fig:5}
\end{figure}

If there are no additional, even approximate conservation laws the 
blocks $U^m|_{{\cal H}_K}$ may be modeled by circular orthogonal 
ensemble (COE) of random matrices for sufficiently large $m$ giving 
the maximal asymptotic (as ${\cal N}\rightarrow\infty$) value of 
relative localization dimension, $\bar{R}_{COE}\approx 0.655$. 
This case corresponds to quantum ergodicity since
$p_{\vec{k}\vec{k}^\prime}(m)$, for $m>M^*$, become pseudo random and 
independent of $\vec{k}$ and $\vec{k}^\prime$, hence the correlation 
function (\ref{eq:corr}) factorizes and yields $C(m)=\ave{J}^2=0$.
Indeed, as we show in fig.4, such behavior is obtained only for 
sufficiently large parameters, say $t=V=4$, 
while for smaller values of parameters $t,V$,
$R(m)$ saturates to a smaller value indicating that there may exist 
approximate conservation laws causing nontrivial localization inside the 
Fock space. Scaling with $1/L$ suggests that even TL of 
$\bar{R}$ is smaller than $\bar{R}_{COE}$ for $t\sim V \sim 1$ (fig.5).

\begin{figure}[htbp]
\begin{center}
\leavevmode
\epsfxsize=3.6in
\epsfbox{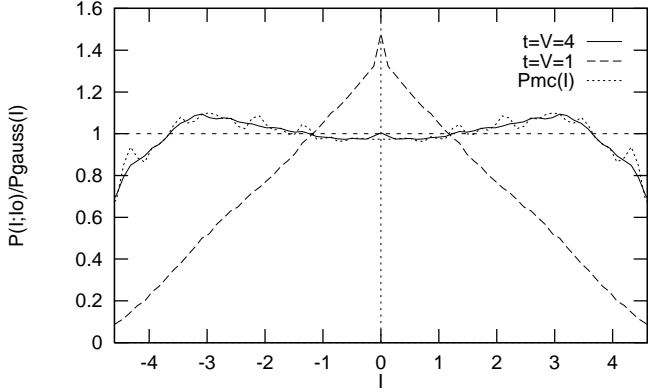}
\end{center}
\caption{Steady-state current distribution divided by a gaussian
$P(I,I_0)/P_{gauss}(I)$ averaged over $279$ initial orbits with 
$|I_0| < 0.08$ in the ergodic, $t=V=4$, and mixed, $t=V=1$, regime, 
and the finite-size microcanonical current distribution 
$P_{mc}(I)$. $L=24,\rho=\frac{1}{4}.$}
\label{fig:6}
\end{figure}

\noindent
Finally, we discuss current fluctuations, or more generaly, current 
distribution $P_\psi(I) = \bra{\psi}\delta(I-J)\ket{\psi}$ giving a 
probability density of having a current $I$ in a state $\ket{\psi}$.
We let the state $\psi$ with a `good' known initial current $I_0$
to evolve for a long time from which we compute a stady-state current 
distribution (SSCD)
$$ P(I;I_0) = \lim\limits_{M\rightarrow\infty}\frac{1}{M}\sum_{m=1}^M
\ave{\delta(I_0-J(0))\delta(I-J(m))}.$$ 
Of course, delta functions should have a finite small width providing
averaging over several states $\ket{\vec{k}}$ with 
$J_{\vec{k}}\approx I_0$.
In the quantum ergodic regime all states eventually become populated,
so SSCD $P(I;I_0)$ should be independent of initial current $I_0$ and 
equal to the {\em microcanonical current distribution}  
$P_{mc}(I) = \ave{\delta(I-J)}$.
It has been shown by elementary calculation that in TL the latter 
becomes a gaussian, $P_{mc}(I) \rightarrow P_{gauss}(I)
=(1/\sqrt{2\pi\ave{J^2}})\exp(-\half I^2/\ave{J^2})$, 
while at any finite size $L$ the first few moments are:
\begin{eqnarray}
\ave{J^2} &=& \frac{N(L-N)}{2(L-1)} \approx \half\rho(1-\rho) L,
\label{eq:mc} \\
\frac{\ave{J^4}}{\ave{J^2}^2} &=& 
\frac{3(L-1)(2N(L-N)-L)}{2N(L-2)(L-N)} \nonumber \\
&=& 3+\frac{3(2\rho(1-\rho)-1)}{2\rho(1-\rho)}\frac{1}{L}+\Ord{L^{-2}}.
\nonumber
\end{eqnarray}
Numerical results for $L=24$ (see fig.6) indicate 
that in the ergodic regime, $t=V=4$, SSCD is already in good agreement with 
microcanonical distribution $P_{mc}(I)$, while in non-ergodic (mixed) 
regime, $t=V=1$, SSCD is {\em localized} on a smaller range indicating that 
the current fluctuation is smaller than $\ave{J^2}$. Note that the mean 
$\bar{I}=\int dI I P(I;I_0)$ is just a persistent current, so
$\bar{I} = \alpha I_0$ (see fig.3).
\\\\
In this Letter we have presented numerical evidence, based on efficiently
coded time evolution of a kicked fermionic system, in support of hypothesis 
that mixed (neither integrable nor ergodic) behavior of a quantum many-body
system may survive TL provided that control parameters are not too far away 
from integrable points. It has been shown that in this regime ideal 
transport is possible. However, if the control parameters are sufficiently 
large we recover quantum ergodycity compatible with random matrix theory 
and normal transport properties. It is interesting to note that at the 
transition point between the two regimes, 
where {\em order parameter} -- {\em stiffness} $D_J|_{L=\infty}$ (inferred 
from $1/L$ scaling) touches zero, the correlation time scale $M^*$ 
drastically increases what is reminiscent of a kind 
of {\em dynamical} phase transition.
Although only data for quarter-filled lattice 
($\rho=\frac{1}{4}$) are presented here, we should stress that essentially
the same conclusions follow from our data for other densities, 
$\rho=\frac{1}{3},\frac{3}{8},\frac{2}{5},\frac{1}{2}$,
with a general rule, that the border of
quantum ergodic regime moves to slightly smaller values of control 
parameters $t,V$ as the density $\rho$ approaches $\frac{1}{2}$.
It should be noted that statistics of eigenphases of evolution operator 
$U$ has been computed as well and it has been found that in the ergodic 
regime level statistics is indeed that of COE while in the mixed regime 
it smoothly interpolates between Poisson and COE.

Discussions with Prof. P. Prelov\v sek, and the financial support by
Ministry of Science and Technology of the Republic of Slovenia are 
gratefully acknowledged.


\begin{references}
\bibitem{footn} Note that the case $t=V=1$ is far beyond 
perturbative treatement of any integrable case.
For example, statistics of eigenphases $\eta_n $ exhibit substantial
level repulsion.
\end{references}
\end{document}